\documentclass[reprint,
superscriptaddress,
amsmath,amssymb,
aip,
]{revtex4-1}

\usepackage{appendix}
\usepackage{graphicx}
\usepackage{dcolumn}
\usepackage{bm}
\usepackage[utf8]{inputenc}
\usepackage[dvipsnames]{xcolor}
\usepackage{enumitem}

\begin{document}
\preprint{AIP/123-QED}

\title{Polar Liquids at Charged Interfaces: A Dipolar Shell Theory}

\author{J. Pedro de Souza}
\affiliation{Department of Chemical Engineering, Massachusetts Institute of Technology, Cambridge, MA, USA}

\author{Alexei A. Kornyshev}
\affiliation{Department of Chemistry and Thomas Young Centre for Theory and Simulation of Materials, Imperial College of London, Molecular Science Research Hub, White City Campus, London W12 0BZ, UK}

\author{Martin Z. Bazant}
\affiliation{Department of Chemical Engineering, Massachusetts Institute of Technology, Cambridge, MA, USA}
\affiliation{Department of Mathematics, Massachusetts Institute of Technology, Cambridge, MA, USA}

\date{\today}

\begin{abstract}
The structure of polar liquids and electrolytic solutions, such as water and aqueous electrolytes, at interfaces underlies numerous phenomena in physics, chemistry, biology, and engineering. In this work, we develop a continuum theory that captures the essential features of dielectric screening by polar liquids at charged interfaces, including oscillations in charge and mass, starting from the molecular properties of the solvent. The theory predicts an anisotropic dielectric tensor of interfacial polar liquids previously studied in molecular dynamics simulations. We explore the effect of the interfacial polar liquid properties on the capacitance of the electrode/electrolyte interface and  on hydration forces between two plane-parallel polarized surfaces. In the linear response approximation, we obtain simple formulas for the characteristic decay lengths of molecular and ionic profiles at the interface.

\end{abstract}

\maketitle

\begin{figure}[b]
\centering
\includegraphics[width=1 \linewidth]{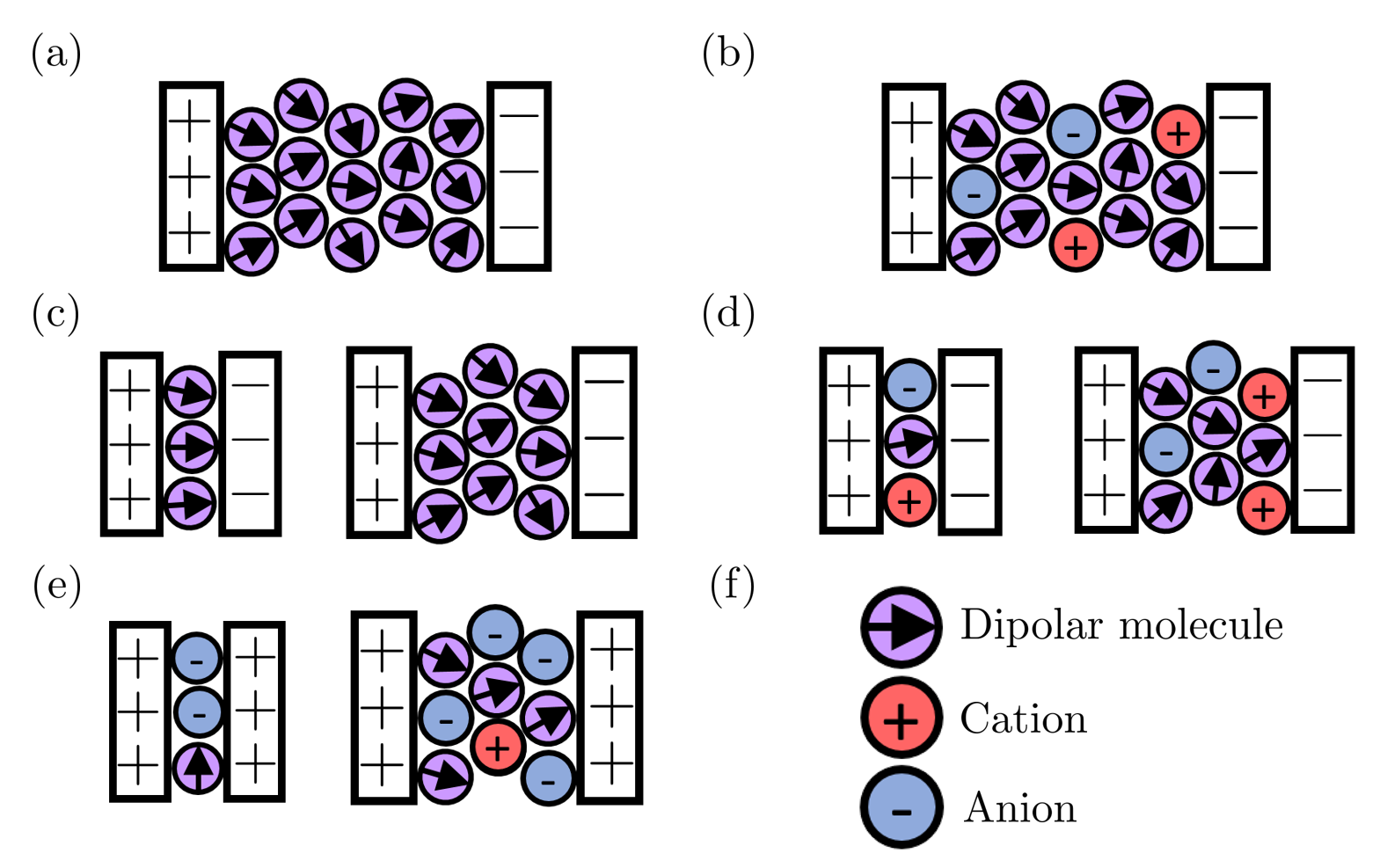}
\caption{Schematic of various systems under consideration in the application of the dipolar shell theory. (a) A pure polar fluid between two oppositely charged surfaces of the same magnitude, obeying overall charge neutrality. (b) A 1:1 electrolyte in a polar solvent with ions of the same size as the dipolar molecules, again with surfaces of equal but opposite charge. (c) A confined pure polar fluid with varying extent of confinement between walls of equal but opposite charge with varying extent of confinement. (d) A 1:1 electrolyte confined between two walls of equal but opposite charge with varying extent of confinement. (e) A 1:1 electrolyte confined between two walls of the same charge with varying extent of confinement. (f) Dipolar molecule, cation, and anion symbols.}
\label{fig:FSchematic}
\end{figure}

\section{Introduction}

 Polar liquids such as water are ubiquitous in all areas of science and engineering, including biological media,\cite{chaplin2006we, ball2008water} electrochemical interfaces,\cite{bard2001fundamentals} colloids,\cite{israelachvili2011intermolecular, lyklema1995solid} synthetic membranes,\cite{geise2014fundamental, marbach2019osmosis} and lubrication. \cite{bowden1950friction, bhushan1995nanotribology}  At charged interfaces, the structure of polar liquids governs the screening of charge by ions in the electrical double layer.\cite{bjorneholm2016water} The key and still not fully understood feature here is the interplay between the molecular structure of the solvent and the correlations in ionic subsystems. Theoretical models of interfacial polar liquids are therefore critical in the design and understanding of electrified interfaces.

Typically, the dielectric properties of interfacial polar liquids are lumped into two main regions of the electrical double layer: (i) the diffuse layer first described by Gouy and Chapman \cite{gouy1910constitution, chapman1913li} in which diffuse ionic charges screen the surface charge  and (ii) the Stern layer composed exclusively of solvent molecules adjacent to the interface,\cite{stern1924theorie} represented as layer of depressed dielectric constant and fixed thickness.\cite{bazant2009towards}  In the diffuse layer, the solvent dielectric constant is usually chosen as its bulk value, and the ions are treated as dilute point charges in the standard Poisson-Boltzmann form.  While such a general approach describes numerous electrochemical measurements, the Gouy-Chapman-Stern (GCS) representation does not capture the microscopic details of the structuring of the fluid near the surface,\cite{kornyshev1982nonlocal, kornyshev1980nonlocal, kornyshev1981nonlocal, bonthuis2013beyond} nor the electric-field dependent response of the polar solvent.\cite{grahame1950effects, abrashkin2007dipolar} One clear shortcoming of the GCS model is its failure to describe oscillatory profiles seen in hydration force measurements \cite{israelachvili1983molecular} and x-ray synchrotron-radiation assessed atomic distribution profiles.\cite{ fenter2004mineral, cheng2001molecular, catalano2011weak}  

A plethora of modifications to the Poisson-Boltzmann theory have been proposed to include the correlated structure and crowding of ions at an interface, usually based on theories of inhomogeneous hard sphere fluids.\cite{bazant2009towards, frydel2012close, gillespie2015review} Applications of such theories to electrolytes usually ignore the solvent molecules by treating the fluid as a constant permittivity medium, $\epsilon$, with hard-sphere ions, the so-called primitive model.\cite{valisko2018systematic, desouza2020continuum, cats2021primitive} While such a model can describe the long-range behavior of dilute electrolytes, it does not capture the short-wavelength structuring of the solvent that affects electrostatic interactions between the ions at a nanometer scale.

In fact, simulations and indirect experimental evidence have demonstrated that the bulk dielectric response of water is nonlocal at short distances. Thus, the studied wave-number dependent static dielectric tensor of water has revealed singularities at short wavelengths, giving rise to the phenomenon of overscreening and alternating bound charges of the polar solvent in response to external perturbation. \cite{bopp1998frequency, bopp1996static, kornyshev1997nonlocal, kornyshev1996shape} At interfaces, molecular dynamics simulations of interfacial water have shown similar overscreening signatures with singularities in the normal component of the anisotropic static dielectric tensor.\cite{bonthuis2011dielectric, bonthuis2012profile, bonthuis2013beyond, schlaich2016water, loche2018breakdown, ruiz2020quantifying}  

To incorporate the dipolar nature of the solvent, a mean-field dipolar Poisson-Boltzmann equation has been developed \cite{abrashkin2007dipolar} and extended further in Refs. \cite{levy2012dielectric, iglic2010excluded, prasanna2013electric, mceldrew2018theory}.  These mean-field dipolar models have not yet captured the overscreening signatures of dipolar molecules at interfaces. Theoretical analysis including the overscreening phenomenon by polar liquids has mainly been limited to situations in which the nonlocal permittivity tensor can be included as an input \cite{kornyshev1983non, belaya1986hydration, levy2020ionic} or through effective Landau-Ginzburg models.\cite{kornyshev1996shape, kornyshev1997nonlocal} These approaches, while generally accurate in comparison to simulations and capable of capturing important features of hydration forces, are not derived from the molecular properties of the solvent and require assumptions to match the bulk screening to the interface.

On the other hand, sophisticated molecular theories including the reference
interaction site model (RISM) can accurately predict the spatially-correlated structures of polar liquids in the bulk and near interfaces.\cite{chandler1972optimized, hirata1982application, pettitt1982integral,ikeguchi1995direct, lue1995application,  beglov1997integral, trokhymchuk1999molecular,  du2000solvation,  fedorov2007unravelling, wu2007density} Due to the complexity of the integral equations involved in solving these theories, some of the analytical tractability and physical transparency is lost in favor of model accuracy, compared to local dipolar theories.\cite{abrashkin2007dipolar} The integral equation theories are therefore difficult to incorporate directly with standard continuum dielectric theory approaches. 

Clearly, a physically-transparent continuum model that incorporates the dipolar, molecular nature of solvent molecules to capture the overscreening behavior at the interface would be useful for understanding the interfacial properties of solvents, solvent mixtures, and electrolytes of  varying ionic composition, including at large applied voltages. 

Here, we derive a modified Langevin-Poisson equation in which we include the nonlocal dielectric response of a polar liquid by employing a weighted-density functional, treating dipolar molecules as shells of charge. The model captures many of the properties of interfacial liquids, including the overscreening of surface charge by the dipolar solvent charges. Singularities in the normal component of the effective static dielectric permittivity at an interface emerge naturally from the theory. After analyzing pure polar liquids, we then include into the theory a finite ion concentration, thus unravelling the fine double layer structure at a charged interface. Further, we apply the theory to describe the hydration forces betweeen two charged surfaces. Finally, we derive a formula for the hydration length, $\lambda_s$,  which depends only on the diameter of the solvent molecule, $d$, and the relative permittivity of the liquid, $\epsilon_r$, where $\lambda_s=d\sqrt{(\epsilon_r-1)/6}$, governing the decay of the oscillations in charge ordering in the polar liquid from a surface. We will explain in the paper the assumptions required to obtain the formula of such extraordinary simplicity, but taken for estimates, it seems to be consistent with existing molecular simulations \cite{bonthuis2011dielectric,bonthuis2012profile, olivieri2021confined, deissenbeck2021dielectric, ruiz2020quantifying}.

In this work, we demonstrate the effects of bulk relative permittivity, surface charge density magnitude from linear to nonlinear response, extent of confinement between two surfaces, and ionic concentration on the interfacial properties of the polar liquid. As depicted in Fig. \ref{fig:FSchematic}, the systems under consideration will include (a) a pure polar liquid between oppositely charged surfaces (b) an electrolyte between two oppositely charged surfaces (c) confined pure polar liquids between oppositely charged surfaces, and (d-e) confined electrolytes between oppositely charged surfaces and identically charged surfaces, respectively. From our theoretical analysis, we demonstrate the importance of the molecular properties of the polar liquid in hydration interactions and double layer capacitance of charged interfaces.


\section{Theory}
 The theoretical approach presented here originates from the model for ionic liquids of ref. \cite{desouza2020interfacial}, to which we add dipolar molecules as hard spheres with dipolar shells of charge. The model treats the equilibrium properties of a concentrated system of dipolar shells in a manner similar to the Langevin-Poisson theories previously described for point dipoles.\cite{abrashkin2007dipolar} For simplicity, we limit our analysis to the case of equally sized ions and dipolar molecules with radius, $R$. This system of ions and dipoles will be positioned between two flat surfaces. We assume that the dipoles and ions in this nanoslit are in equilibrium with a reservoir with fixed bulk concentrations of both ions and dipoles, within the grand canonical ensemble.  
 
\subsection{Density functional}

The theoretical framework is based on a definition of the Helmholtz free energy functional of the system, and could be classified as a classical Density Functional Theory approach. The Helmholtz free energy, $\mathcal{F}$ can be split into three parts: an ideal part $\mathcal{F}^{\mathrm{id}}$, an excess part accounting for excluded volume effects, $\mathcal{F}^{\mathrm{ex}}$, and an electrostatic part, $\mathcal{F}^{\mathrm{el}}$, 
\begin{equation}
    \mathcal{F}=\mathcal{F}^{\mathrm{id}}+\mathcal{F}^\mathrm{ex}+\mathcal{F}^\mathrm{el}.
\end{equation}

As a standard definition, the ideal part of the free energy density is given by:
\begin{equation} \label{eq:eq_Fid}
\mathcal{F}^{\mathrm{id}}[\{c_i(\mathbf{r})\}]=\sum_i  k_{B}T \int d\mathbf{r}\,c_i(\mathbf{r})\big[\ln (\Lambda_i^3 c_i(\mathbf{r})) - 1\big],   
\end{equation}
where $k_B$ is the Boltzmann constant, $T$ is the absolute temperature, $c_i$ is the concentration of species $i$, and $\Lambda_i$ is the de Broglie wavelength for species $i$. The ideal part of the chemical potential for species $i$, $\mu_i^\mathrm{id}$, relative to some reference bulk solution denoted as $b$, is thus:
\begin{equation}
    \beta\mu_i^\mathrm{id}=\beta\left(\frac{\delta \mathcal{F}^\mathrm{id}}{\delta c_i}-\frac{\delta \mathcal{F}^\mathrm{id}}{\delta c_i}\Big\rvert_b\right)=\ln\left(\frac{c_i}{c_{i0}}\right),
\end{equation}
where $c_{i0}$ is the concentration in the bulk and $\beta$ is the inverse thermal energy, $\beta^{-1}=k_B T$.

For the excess free energy density, we will assume that all species are approximately spherical and equal in size. Here, we adopt a weighted-density approximation from ref. \cite{desouza2020interfacial} that was constructed to recover the Carnahan-Starling equation of state: 
 \begin{equation}\label{eq:eq_Fex}
\mathcal{F}^\mathrm{ex}[\bar{c}_i(\mathbf{r})]=\frac{k_{B}T}{v}\int d\mathbf{r}\,\bigg[\dfrac{1}{1 - \bar{\eta}} - 3\bar{\eta} + \dfrac{1}{(1 - \bar{\eta})^{2}}\bigg].
\end{equation}
Here, $v$ is the volume of a molecule, $\eta=\sum_i v c_i$ is the local filling fraction, and $\eta_0=\sum_i v c_{i0} $ is the bulk filling fraction. The bar notation denotes a convolution with the volumetric weighting function, $\bar{\eta}=w_v*\eta$, where $w_v(r)=\Theta(R-\mid r\mid )/v$, is a Heaviside step function that only turns on within the volume of the sphere, with $R$ defined as the radius of the molecular sphere. The $*$ operator corresponds to a convolution integral, $f*g=\int d\mathbf{r^\prime} f(\mathbf{r^\prime})g(\mathbf{r-r^\prime})$. Therefore, the excluded volume interactions appear in a non-local fashion in the chemical potential, describing the filling within a molecular-sized neighborhood of a point. The associated weighted excess chemical potential, $\bar{\mu}_i^\mathrm{ex}$,is:
\begin{equation}
\begin{split}
        \beta\bar{\mu}_i^\mathrm{ex}&=\beta\left(\frac{\delta \mathcal{F}^\mathrm{ex}}{\delta c_i}-\frac{\delta \mathcal{F}^\mathrm{ex}}{\delta c_i}\Big\rvert_b\right)\\&=w_v *\left(\frac{8\bar{\eta}-9\bar{\eta}^2+3\bar{\eta}^3}{(1-\bar{\eta})^3}-\frac{8 \eta_0-9\eta_0^2+3\eta_0^3 }{(1-\eta_0)^3}\right).
\end{split}
\end{equation}

\subsection{Derivation of electrostatic variables}
The key development in the theory presented in this work is the electrostatic part of the free energy. Here, we spread the ionic charge and bound charges on dipoles over their surface, so that they act electrostatically as shells rather than points. The smeared shell charge appears in the mean-field Poisson equation for both the ions and the polar liquid molecules. The charged shell formulation here is directly based on the theory for concentrated ionic liquids presented in ref. \cite{desouza2020interfacial}. The approach is preceded by similar theoretical models for electrolytes composed of ions with intramolecular charge distributions,\cite{frydel2013double, wang2010fluctuation, may2008bridging, frydel2014mean, frydel2016double, adar2019screening} as well as charged shell representation of the mean-spherical approximation.\cite{blum1991relation, roth2016shells, jiang2021revisiting} The charged shell approximation is applicable not only to ions and dipoles with charge form factors in the shape of a spherical shell, but also for hard sphere ions and dipoles with point charges and point dipole moments at their centers in which the electrostatic potential can only develop beyond the ionic radius or dipolar molecule radius.  The electrostatic part of the free energy, defined in terms of weighted densities, is therefore:
\begin{equation}\label{eq:eqFel_main}
    \mathcal{F}^\mathrm{el}[ \phi, \bar{\rho}_e, \mathbf{\bar{P}}] =  \int d\mathbf{r} \Big\{-\frac{\epsilon_0}{2}(\nabla\phi)^2+ \bar{\rho}_e \phi+\mathbf{\bar{P}}\cdot \nabla\phi\Big\},
\end{equation}
 where $\phi$ is the electrostatic potential, $\epsilon_0$ is the permittivity of free space, $\bar{\rho}_e=w_s*\rho_e$ is the weighted charge density, and $\bar{\mathbf{P}}=w_s*\mathbf{P}$ is the weighted polarization vector, originating from the weighted bound charge on the dipolar molecules, $\bar{\rho}_b=w_s*\rho_b=\nabla \cdot \mathbf{\bar{P}} $. Here, the convolution with the weighting function $w_s(r)=\delta(R-\mid r \mid)/(4\pi R^2)$ homogenizes the charge and polarization over a spherical shell.

Using the definition of the polarization vector, $\mathbf{P}$, as the concentration of dipoles, $c_w$, multiplied by their individual dipole moments, $\mathbf{p}$, we can next define the weighted polarization vector, $\bar{\mathbf{P}}$: 
\begin{equation}
  \mathbf{P}=c_w \mathbf{p}, \quad \mathbf{\bar{P}}=w_s *\left(c_w \mathbf{p}\right),   
\end{equation}
for which there will be a distribution of dipole orientations that we will ultimately need to average over.
The electrochemical potential of the dipole, $\mu_w$ can be found by taking the variational derivative of the free energy with respect to the dipole concentration:
\begin{align}
    \beta\mu_w=\ln\left(\frac{c_w}{c_{w0}}\right)+\beta \mathbf{p}\cdot\nabla \bar{\phi}+\beta\bar{\mu}^\mathrm{ex}_w.
\end{align}
Here, $\bar{\phi}=w_s *\phi$ is the weighted electrostatic potential, and $\bar{\mu}^\mathrm{ex}_w=w_v*{\mu}^\mathrm{ex}_w$ is the weighted excess chemical potential. Mathematically, the weighted electrostatic potential and weighted excess chemical potential emerge due to the minimization of the free energy with respect to the concentration variables, which are present in the free energy in terms of convolutions with weighting functions. Physically, these are a result of the delocalization of bound charge over the dipolar molecule surface and the nonlocal packing effects, respectively. This procedure embeds the finite size of dipolar molecules into the theory, which plays a key role in capturing the effects of layering and decoupling the packing periodicity from the longer range electrostatic correlations. Thus, although this approach should still be classified as a mean field theory, it makes an essential step towards accounting for molecular structure of the liquid.    The dipole concentration at a given position is thus:
\begin{equation}
    c_w=c_{w0}\exp\left(-\beta\mathbf{p}\cdot \nabla \bar{\phi} -\beta\bar{\mu}^\mathrm{ex}_w\right).
\end{equation}
Therefore, the local dipole concentration depends on the angle, $\theta$, between the dipole moment of the molecule and the weighted electric field, $\overline{\mathbf{E}}=w_s*\mathbf{E}=-\nabla \bar{\phi}$, where
\begin{equation}
   \mathbf{p}\cdot \nabla \bar{\phi}=p_0 \mid\nabla \bar{\phi}\mid \cos\theta, 
\end{equation}
assuming a constant effective dipole moment magnitude, $p_0$. Averaging over the possible orientations of the molecule gives:
\begin{equation}
\begin{split}
        \langle c_w\rangle=c_{w0}e^{ -\beta\bar{\mu}^\mathrm{ex}_w}\left\langle e^{-\beta p_0\mid \nabla \bar{\phi}\mid \cos\theta} \right\rangle
\end{split}
\end{equation}
for the dipole concentration and:
\begin{equation}
    \langle \mathbf{P}\rangle=c_{w0} p_0 e^{ -\beta\bar{\mu}^\mathrm{ex}_w}\frac{\nabla \bar{\phi}}{\mid\nabla \bar \phi \mid}\left\langle \cos(\theta) e^{-\beta p_0\mid \nabla \bar{\phi}\mid \cos\theta} \right\rangle
\end{equation}
for the polarization vector, where $\langle \rangle$ denotes the average over $\theta$. Averaging over the dipole orientations gives:
\begin{equation}
\begin{split}
     \left\langle e^{-\beta p_0\mid \nabla \bar{\phi}\mid \cos\theta} \right\rangle&=\frac{1}{2}\int_0^{\pi} e^{-\beta p_0\mid \nabla \bar{\phi}\mid \cos\theta} \sin \theta d\theta \\
     &=\frac{\sinh\left(\beta p_0 \mid\nabla\bar{\phi}\mid\right)}{\beta p_0 \mid\nabla\bar{\phi}\mid}
\end{split}
\end{equation}
and
\begin{equation}
\begin{split}
     \left\langle \cos(\theta)  e^{-\beta p_0\mid \nabla \bar{\phi}\mid \cos\theta} \right\rangle&=\frac{1}{2}\int_0^{\pi} e^{-\beta p_0\mid \nabla \bar{\phi}\mid \cos\theta} \cos\theta \sin \theta d\theta \\
     &=-\frac{\sinh\left(\beta p_0 \mid\nabla\bar{\phi}\mid\right)}{\beta p_0 \mid\nabla\bar{\phi}\mid}\mathcal{L}\left(\beta p_0 \mid\nabla\bar{\phi}\mid\right)
\end{split}
\end{equation}
where $\mathcal{L}(x)=\coth(x)-1/x$ is the Langevin function. Therefore, the dipole concentration can be written as:
\begin{equation}
     \langle c_w\rangle=c_{w0}e^{ -\beta\bar{\mu}^\mathrm{ex}_w}\frac{\sinh\left(\beta p_0 \mid\nabla\bar{\phi}\mid\right)}{\beta p_0 \mid\nabla\bar{\phi}\mid},
\end{equation}
and the polarization density can be expressed as:
\begin{equation}
    \langle \mathbf{P}\rangle= -p_0 \langle c_w\rangle \frac{\nabla \bar{\phi}}{\mid\nabla \bar \phi \mid}\mathcal{L}\left(\beta p_0 \mid\nabla\bar{\phi}\mid\right).
\end{equation}
In turn, the weighted polarization vector is defined as:
\begin{align}\label{eq:eqP_bar_def}
    \langle \mathbf{\bar{P}}\rangle= -w_s*\left[p_0 \langle c_w\rangle\frac{\nabla \bar{\phi}}{\mid\nabla \bar \phi \mid} \mathcal{L}\left(\beta p_0 \mid\nabla\bar{\phi}\mid\right)\right].
\end{align}
Moving forward, we drop the bracket notation, such that $\mathbf{P}$ refers to $\langle \mathbf{P}\rangle$, $\mathbf{\bar{P}}$ refers to $\langle \mathbf{\bar{P}}\rangle$, and $c_w$ refers to $\langle c_w \rangle$. 
 
Along with the description of the polarization vector, we must also describe the ionic charge when there is a non-zero electrolyte concentration. The ionic electrochemical potential is defined as:
\begin{equation}
    \beta\mu_i=\ln\left(\frac{c_i}{c_{i0}}\right)+z_i e\beta \bar{\phi}+\beta\bar{\mu}^\mathrm{ex}_i.
\end{equation}
Therefore, the distributions of the ions are given by:
\begin{align}
    c_i=c_{i0}\exp\left(-\beta e\bar{\phi}-\beta\bar{\mu}^\mathrm{ex}_i\right).
\end{align}
For a 1:1 solution of concentration $c_0$, the electrolyte charge density is therefore:
\begin{equation}\label{eq:eqIonicCharge}
    \rho_e=-2e c_0\sinh\left(\beta e \bar{\phi}\right)e^{-\beta\bar{\mu}^\mathrm{ex}_i},
\end{equation}
and the weighted electrolyte charge density is $\bar{\rho}_e=w_s* \rho_e$.

 The Poisson equation consistent with the free energy density in Eq. \ref{eq:eqFel_main} is:
\begin{equation}\label{eq:eq_A}
    -\epsilon_0 \nabla^2\phi=-\nabla\cdot\mathbf{\bar{P}}+\bar{\rho}_e.
\end{equation}
as shown in Appendix A. The source of the electric field is therefore the delocalized shells of charge of the ions and bound charge on dipolar molecules, mathematically appearing as convolutions with the weighting function, $w_s$. Specifically, the overall charge density in the Poisson equation includes both the weighted bound charge density on the dipolar molecules, $\bar{\rho}_b=-\nabla\cdot\mathbf{\bar{P}}$, and the weighted ionic charge, $\bar{\rho}_e$.

\begin{figure*}
\centering
\includegraphics[width=1 \linewidth]{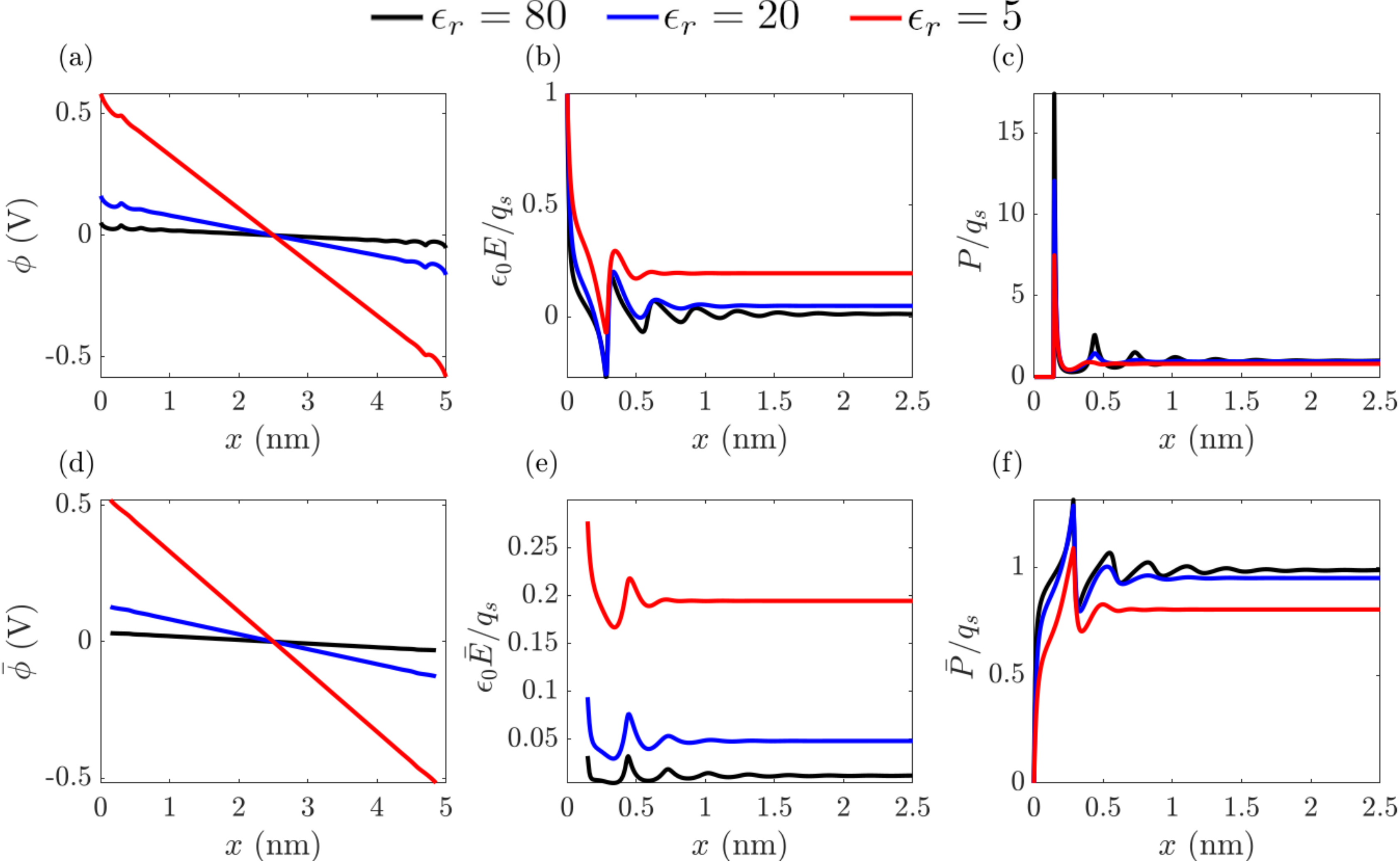}
\caption{Electrostatic screening by pure polar liquids between two surfaces of opposite charge shown for different values of the bulk dielectric constant, comparing the weighted and nonweighted quantities. The curves are generated by solving Eq. \ref{eq:1D_form} with $\bar{\rho}_e=0$. Variables are plotted as functions of the normal coordinate, $x$, zooming into the profiles emerging from the left interface for b-c and e-f. The selected bulk dielectric permittivities, correspond to values of $p_0=4.86$ D for $\epsilon_r=80$, $p_0=2.38 $ D for $\epsilon_r=20$, and $p_0=1.09 $ D for $\epsilon_r=5$, keeping all other parameters constant ($T=300$ K, $L=5$ nm, $c_{w0}=55$ M, $d=0.285$ nm, and $q_s=0.01$ C/m$^2$). (a) Electrostatic potential, $\phi$, (b) electric field, $E=-\phi^\prime$, (c) polarization density, $P$, (d) weighted electrostatic potential, $\bar{\phi}$,  (e) weighted electric field, $\bar{E}=-\bar{\phi}^\prime$, and (f)  and weighted polarization density, $\bar{P}$. The local variables in a-c describe the ``measured" local electrostatic response of the system, while the weighted potential and weighted electric field in d-e determine the electrochemical potential and orientation of dipoles. The weighted polarization vector in f corresponds to the polarization arising from the delocalized bound charge on the dipolar shells.  }
\label{fig:F_eps_var1}
\end{figure*}

\subsection{Reducing to one dimension}
 In the 1D geometry between two flat plates, the weighting function formulas must be modified~\cite{roth2010fundamental} to integrate over the $y$ and $z$ dimensions, since all variables depend only on $x$. Physically, the spherical shell of charge corresponding to $w_s$ becomes equivalent to a line of charge with length $2R$ and uniform charge per length (the differential area of a sphere per differential in the axial coordinate). The spherical Heaviside weighting function, $w_v$, becomes a quadratic function, corresponding to the differential volume of a sphere per differential in the axial coordinate. Their modified forms are:
\begin{align}
    &w_v(x-x^\prime)=\frac{\pi\left(R^2-(x-x^\prime)^2\right)}{v}\Theta\left(R-\mid x-x^\prime\mid\right)\\
    &w_s(x-x^\prime)=\frac{1}{2 R}\Theta\left(R-\mid x-x^\prime\mid\right).
\end{align}

In 1D, the integro-differential equation for the electrostatic potential is:
\begin{equation}\label{eq:1D_form}
    0=\epsilon_0 \frac{d^2\phi}{dx^2}-\frac{d\bar{P}}{dx}+\bar{\rho}_e
\end{equation}
where
\begin{equation}
    {P}=-p_0 c_{w0}e^{ -\beta\bar{\mu}^\mathrm{ex}_w}\frac{\sinh\left(\beta p_0 \bar{\phi}^\prime\right)}{\beta p_0 \bar{\phi}^\prime}\mathcal{L}\left(\beta p_0 \bar{\phi}^\prime\right),
\end{equation}
and $\bar{\phi}^\prime=d\bar{\phi}/dx$ denotes a derivative with respect to $x$. In this geometry, the average orientation of the dipoles relative to the $x$-axis can be expressed as:
\begin{align}
    \langle \cos(\theta) \rangle=- \mathcal{L}\left(\beta p_0 \bar{\phi}^\prime\right).
\end{align}

We assume that the surface charge is uniformly distributed on bounding flat hard walls at $x=0$ and $x=L$. For simplicity, here, the surface charge density is assumed not to have any finite size, so the boundary conditions reduce to:
\begin{equation}\label{eq:eqBCs}
\begin{split}
        &\left(-\epsilon_0\phi^\prime\right)\Big \rvert_{x=0}=  q_s \\
        &\left(-\epsilon_0\phi^\prime\right)\Big \rvert_{x=L}= \pm q_s 
\end{split}
\end{equation}
where we have the surface charge density of magnitude $q_s$ on each side. Depending on the scenario under investigation, the charge on each surface is either opposite in sign or the same in sign, as sketched in Fig. \ref{fig:FSchematic}. If the charge at x=L is negative, we choose ``+" in the second line of Eq. \ref{eq:eqBCs}, and the opposite is true if the charge is positive. The local dipolar and ionic concentrations are zero in the regions $x<R$ and $x>L-R$ owing to the hard sphere repulsion with the flat bounding surfaces. To solve these equations, we discretized them using finite difference formulas. 

\begin{figure*}
\centering
\includegraphics[width=1 \linewidth]{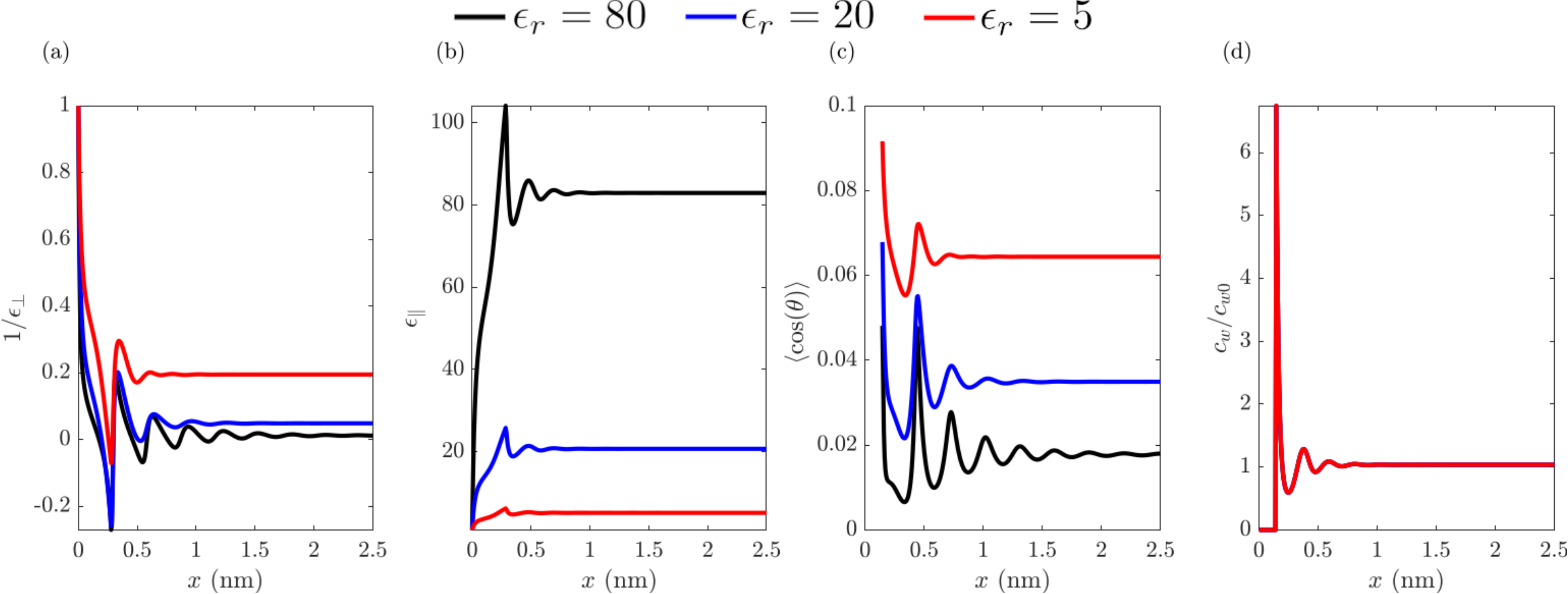}
\caption{Dielectric tensor, average orientation, and density of pure polar fluids between two surfaces of opposite charge, generated for model polar liquids with different values of the bulk dielectric constant. The curves are generated by solving Eq. \ref{eq:1D_form} with $\bar{\rho}_e=0$.
Variables are plotted as functions of the normal coordinate, $x$, zooming into the profiles emerging from the left interface. The results are plotted for three pure polar fluids, corresponding to $p_0=4.86$ D for $\epsilon_r=80$, $p_0=2.38 $ D for $\epsilon_r=20$, and $p_0=1.09 $ D for $\epsilon_r=5$, keeping all other parameters constant ($T=300$ K, $L=5$ nm, $c_{w0}=55$ M, $d=0.285$ nm, and $q_s=0.01$ C/m$^2$). (a) Normal component of the dielectric tensor, ${\epsilon}_\perp$, plotted in terms of its inverse. (b) Tangential component of the dielectric tensor, ${\epsilon}_\parallel$. (c) Average orientation of dipolar molecules, $\langle \cos(\theta)\rangle$. (d) Density profile of dipolar molecules, $c_w$, normalized by the bulk value. In (d), the density profiles are closely overlapping each other, due to the low applied surface charge.  }
\label{fig:F_eps_var2}
\end{figure*}

\subsection{Extracting the effective local dielectric tensor}
\subsubsection{Normal component}
A fundamental calculation involves extracting the static dielectric profile from the predicted polarization vector. Here, we choose a permittivity definition that is consistent with the weighted Poisson equation. The self-consistent definition of the normal permittivity is given by:
\begin{equation}
    {\epsilon}_\perp=1-\frac{\bar{P}}{\epsilon_0\phi^\prime}
\end{equation}
which represents the total displacement vector, $D=-\epsilon_0\phi^\prime+\bar{P}$, divided by the electric field, both extracted directly from our model. Note that while the electrochemical potential of the dipoles depends on the weighted electrostatic field, the displacement vector includes a contribution from the local electric field ($-\phi^\prime$) and also the weighted polarization vector ($\bar{P}$).  The modified Poisson equation can be written in terms of the effective normal permittivity in the form:
\begin{align}\label{eq:perp_perm}
    \frac{d}{dx}\left(\epsilon_0{\epsilon}_\perp\frac{d\phi}{dx}\right)=-\bar{\rho}_e.
\end{align}
Here, the effective normal permittivity includes only the polarization of the solvent, and does not include the polarization from the ions. We will return to the nuances of the definition of the normal permittivity in the analysis of concentrated electrolytes.

\subsubsection{Tangential component}

While we assume no tangential component of the field in the solution of our model, we can also use the model to quantify the extent of tangential polarizability of the interfacial polar liquid in response to macroscopic electric fields tangential to the plane of the interface. If the tangential electric field is constant and weak relative to the normal electric field, then we can extract it as a small, constant perturbation upon the normal field. For example, we can assume a small perturbative component of the electric field in the $y$ direction, $E_y$, that satisfies $\mid E_y\mid \ll \mid E_x\mid$. We can approximate the magnitude of the gradient of the weighted electrostatic potential  as: $\mid \nabla \bar{\phi}\mid\approx \mid \bar{E}_x \mid=\mid \bar{\phi}^\prime\mid $, where the prime notation still refers to derivatives in the $x$-direction, and $\bar{E}_x$ is the weighted electric field in the $x$-direction. If we apply such an assumption to the $y$-component of Eq. \ref{eq:eqP_bar_def}, the displacement vector in the $y$-direction is therefore:
\begin{align}
\begin{split}
        D_y&=\epsilon_0 E_y+\bar{P}_y\approx \epsilon_0{E}_y+w_s*\left[p_0 c_w \frac{\bar{E}_y}{\mid \bar{E}_x\mid}\mathcal{L}\left(\beta p_0 \mid \bar{E}_x\mid \right)\right]\\
        &\approx\epsilon_0{E}_y+w_s*\left[p_0 c_w \frac{\bar{E}_y}{ \bar{\phi}^\prime}\mathcal{L}\left(\beta p_0 \bar{\phi}^\prime \right)\right].
\end{split}
\end{align}

Next, we divide the tangential ($y$) component of the displacement vector by the tangential electric field ($E_y$). Since we assume the tangential electric field is constant due to the system's translational invariance in the $yz-$plane, it can be treated as a constant for the differentiation or convolution operations, so that $E_y\approx \bar{E}_y$. Through this process, the tangential permittivity, ${\epsilon}_\parallel$, can be defined as:
\begin{align}
    {\epsilon}_\parallel\approx1-w_s * \left(\frac{P}{\epsilon_0\bar{\phi}^\prime}\right).
\end{align}

\begin{figure*}
\centering
\includegraphics[width=1 \linewidth]{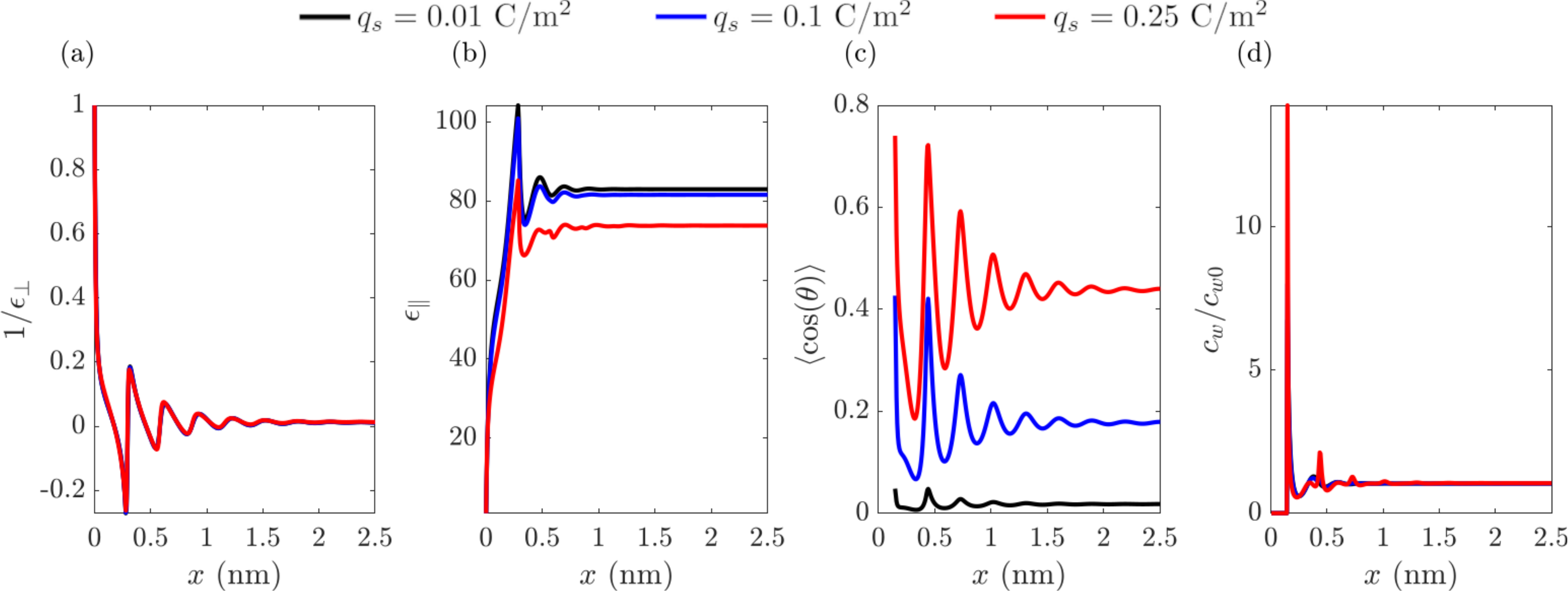}
\caption{Dielectric tensor, average orientation, and density of pure polar fluids between two surfaces of opposite charge for varying surface charge density. The curves are generated by solving Eq. \ref{eq:1D_form} with $\bar{\rho}_e=0$.
Variables are plotted as functions of the normal coordinate, $x$, zooming into the profiles emerging from the left interface. The results are plotted for varying surface charge density ($q_s=0.01$ C/m$^2$, $q_s=0.1$ C/m$^2$, and $q_s=0.25$ C/m$^2$), keeping all other parameters constant ($T=300$ K, $L=5$ nm, $c_{w0}=55$ M, $d=0.285$ nm, and $p_0=4.86$ D). (a) Normal component of the dielectric tensor, ${\epsilon}_\perp$, plotted in terms of its inverse. (b) Tangential component of the dielectric tensor, ${\epsilon}_\parallel$. (c) Average orientation of dipolar molecules, $\langle \cos(\theta)\rangle$. (d) Density profile of dipolar molecules, $c_w$, normalized by the bulk value.  }
\label{fig:F_qs_var2}
\end{figure*}

\section{Results}

The equations are solved between two surfaces with fixed charge densities of magnitude $q_s$ and separation distance $L$. The baseline parameters correspond to the effective values for water: dipolar molecule concentration $c_{w0}=55$ M (corresponding to $\approx$33 molecules per nm$^3$), temperature $T=300$ K, and a diameter of $d=0.285$ nm. While $\epsilon_r$ is the bulk dielectric constant far from the interface, the local static dielectric tensor can vary as a function of position. In the bulk, the relative permittivity as given by the dipolar model for small perturbations is given by:\cite{abrashkin2007dipolar}
\begin{align}
    \epsilon_r=1+\frac{\beta c_{w0} p_0^2}{3\epsilon_0}.
\end{align}
 For a dipolar concentration of $c_{w0}=55$ M at $T=300$ K, the bulk dielectric constant of $\epsilon_r=80$ requires an effective dipole moment of $p_0=4.86$ D. This effective value is significantly larger than the actual dipole moment of water, 1.8 D. The effective dipole moment accounts for correlations between the orientation of a single dipolar molecule and the orientation of its nearest neighbors, as accounted for in more sophisticated bulk dielectric theories.\cite{booth1951dielectric, kirkwood1939dielectric, gongadze2013spatial, levy2012dielectric} Here, we lump these effects into the effective dipole moment in our model, similar to previous dipolar Poisson-Boltzmann approaches.\cite{abrashkin2007dipolar, mceldrew2018theory} The default separation distance between the two charged surfaces is $L=5$ nm, and the default surface charge density magnitude is $q_s=0.01$ C/m$^2$. 

With the inclusion of the polar fluid, the parameter space for the system under investigation is large. We therefore divide our results into five parts: (A) First, we present results for pure polar fluids between two opposite surfaces. We formally investigate how the interfacial electrostatic properties change with varying bulk dielectric constant, and also how nonlinear saturation of the dipole orientation arises at high surface charge. We also show the complicated layering of molecular orientation for dipolar fluids confined to the sub-nanometer scale.   (B) Second, the interfacial electrostatic properties are investigated with a non-zero ionic concentration between oppositely charged surfaces. Here, the results are presented for varying ionic concentrations and for varying surface charge magnitudes. (C) Next, the theory is applied to understanding hydration interactions between two surfaces of (i) opposite charge with and without ions present and (ii) the same charge with a non-zero ionic concentration. (D) The double layer capacitance with ions present is then investigated, ensuring non-overlapping double layers with large separation distances between the surfaces. (E) Finally, the equations are linearized and cast into a differential form, which gives analytical decay lengths describing the layering of charge and mass at the interface.

\subsection{Pure polar fluids: Interfacial dielectric structure}
The system of a pure polar fluid is rare in practice, but it is a useful reference system to showcase the dipolar shell theory predictions. In order to maintain electroneutrality, the bounding surfaces must have equal but opposite charge density, since the pure polar fluid does not have any net charge. Within this system, we will highlight the influence of the fluid bulk permittivity, $\epsilon_r$, the strength of the electric field in the system that is set by the surface charge density on the boundaries, $q_s$, and the confinement extent of the fluid given by the surface separation distance, $L$.

To start, we investigate the effect of the effective dipole moment, $p_0$, on the resulting potential and electric field distribution. Since $p_0$, of course, determines the bulk macroscopic dielectric constant, $\epsilon_r$, we may say that we will trace the effect of the latter on the local properties near the interface, although the variation $\epsilon_r$ is itself the result of variation of $p_0$. 
 In order to study fluids of different bulk dielectric constants, we vary $p_0$ to take on the values $p_0=2.36$ D for $\epsilon_r=20$ and $p_0=1.09$ D for $\epsilon_r=5$, keeping all other variable constant. 

In Fig. \ref{fig:F_eps_var1}, the electrostatic potential, electric field, and polarization density are plotted along with their weighted counterparts for fluids of varying bulk permittivity, $\epsilon_r$.  For each fluid, the electrostatic potential, $\phi$, oscillates near the surface, within the first nanometer. The potential difference across the nanoslit is greatest for the least polar fluid with $\epsilon_r=5$, since dielectric screening is the weakest for this system. The oscillations in the potential lead to sharp cusps and oscillations in the electric field, $E$, for all the fluids. The electric field even reverses signs at some points, corresponding to reversal in the local electric field direction, corresponding to \textit{overscreening} of the surface charge. The oscillations and sign reversals in the electric field are stronger for the more polar fluids with higher $\epsilon_r$. The local polarization density, $P$, has related signatures, where the local polarization density magnitude exceeds the imposed displacement field magnitude set by the surface charge density on the bounding walls.   For the weighted variables, $\bar{\phi}$, $\bar{E}$, and $\bar{P}$, the weighting operation smooths out the oscillations compared to the local variables, but does not eliminate them. 

The weighted electrostatic potential, $\bar{\phi}$, and weighted electric field, $\bar{E}$, determine the local electrostatic energy and orientation of the dipolar shells in the theory. The weighted polarization density, $\bar{P}$, contributes to the overall displacement field, $D=\epsilon_0 E+\bar{P}$. Therefore, the overscreening of the surface charge occurs when $\bar{P}$ exceeds $q_s$, where the cumulative bound dipolar shell charge exceeds the surface charge density. All the fluids experience at least one overscreening peak in the weighted polarization profile. However, the oscillations decay more rapidly and the overscreening peaks are smaller for the lower bulk permittivity liquids. From these profiles, we can deduce that the overscreening phenomenon is essential to describing the structuring of polar liquids with large bulk dielectric constant, and is microscopically sensitive to the effective dipole moment, $p_0$.

\begin{figure}
\centering
\includegraphics[width=1 \linewidth]{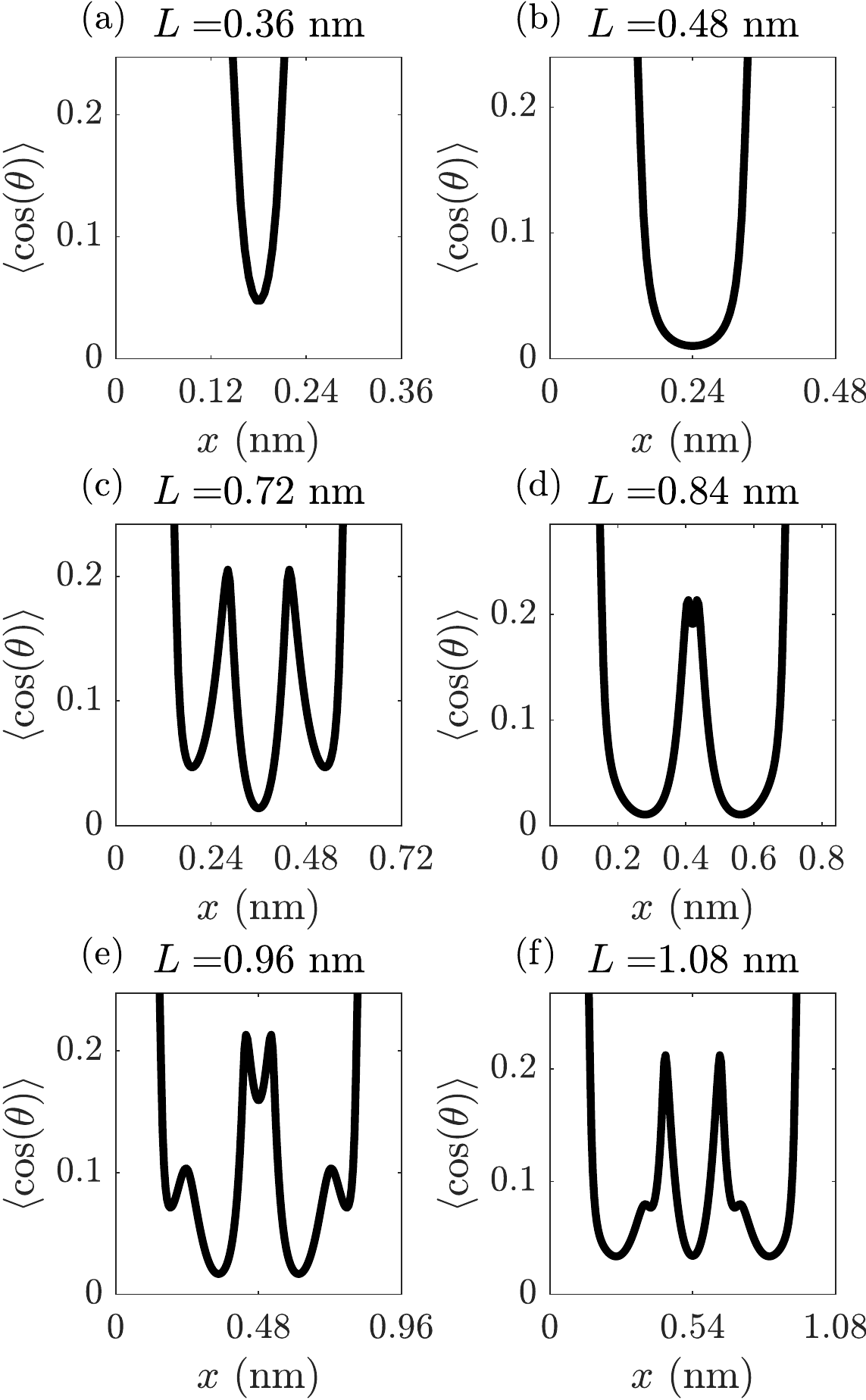}
\caption{Orientation of molecules in a pure polar fluid between two oppositely charged surfaces as a function of the confinement distance between the surfaces. The curves are generated by solving Eq. \ref{eq:1D_form} with $\bar{\rho}_e=0$. The results are plotted for indicated separation distances between the two confining charged surfaces, $L$, keeping all other parameters constant ($T=300$ K, $c_{w0}=55$ M, $d=0.285$ nm, $p_0=4.86$ D, $q_s=0.05$ C/m$^2$).}
\label{fig:F_cos_theta_L}
\end{figure}

\begin{figure*}
\centering
\includegraphics[width=1 \linewidth]{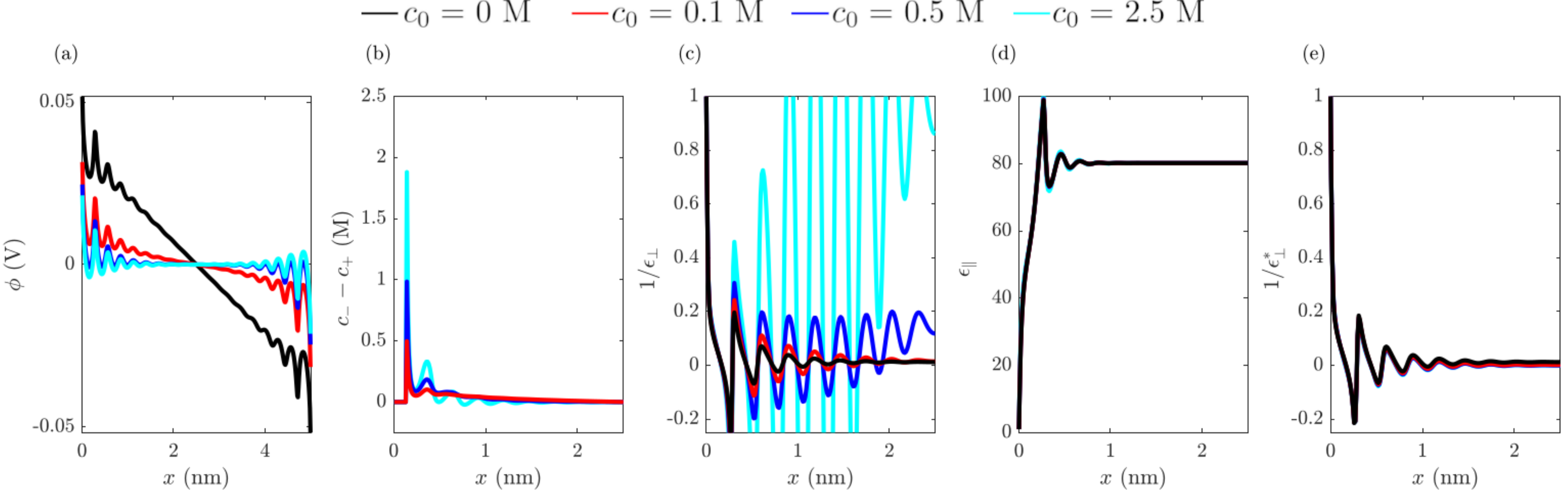}
\caption{Electrolyte screening behavior between two surfaces of opposite charge for varying ionic concentration. The curves are generated by solving Eq. \ref{eq:1D_form} with $\bar{\rho}_e\neq 0$. The ionic concentration is varied between $c_0=0$ M, $c_0=0.1$ M, $c_0=0.5$ M, and $c_0=2.5$ M, keeping all other parameters constant ($T=300$ K, $L=5$ nm, $c_{w0}=55$ M, $d=0.285$ nm, $q_s=0.01$ C/m$^2$, and $p_0=4.86$ D). (a) Electrostatic potential, $\phi$. (b) Local difference in ionic concentration, porportional to the local charge density. (c) Normal component of the dielectric tensor (effective--accounting only for the polarization of solvent), ${\epsilon}_\perp$, plotted in terms of its inverse. (d) Tangential component of the dielectric tensor, ${\epsilon}_\parallel$. (e) Normal component of the general dielectric tensor, ${\epsilon}_\perp^*$ (accounting for the polarization of the solvent and ions).  }
\label{fig:F_conc_var}
\end{figure*}

Next, we can use the electrostatic variables to determine the components of the dielectric tensor near the interface, as shown in Fig. \ref{fig:F_eps_var2}, for fluids with different bulk dielectric constants. Due to the weighted polarization density overscreening the surface charge density, the normal component of the dielectric tensor has singularities. The tangential component, on the other hand, does not have the same overscreening structure since it is set by a long range tangential electric field, and the tangential component of the dielectric function varies closely with the local dipole concentration, $c_w$. The orientation of the dipoles, $\langle \cos(\theta)\rangle$, is actually higher for the fluid that is the least polar.  This fact appears because the least polar fluid with $\epsilon_r=5$ corresponds to the weakest dielectric screening of the electric field. Even though large differences are observed in the dielectric profiles, at the low charge density of $q_s=0.01$ C/m$^2$, the dipole concentration is not very strongly affected by electrostatics, as shown in Fig. \ref{fig:F_eps_var2}(d). Instead, the dipole density is dominated by the packing effects embedded in $\bar{\mu}^\mathrm{ex}_i$, which is independent of electrostatics at small potentials. The large contact value for the density is also governed by the nonlocal packing effects, similar to uncharged hard-sphere fluids. In turn, the excess chemical potential owing to packing plays a minor role in the overscreening structure in the normal component of the dielectric tensor. The overscreening signatures can therefore be attributed to the delocalization of the bound charge on dipoles over the dipole molecule surface. The remarkable anisotropic static dielectric tensor predicted by the dipolar shell theory here is similar to the reported Molecular Dynamics simulations of the dielectric properties of interfacial water.\cite{bonthuis2012profile, bonthuis2013beyond, olivieri2021confined, deissenbeck2021dielectric, ruiz2020quantifying} The overscreening signatures also qualitatively match the results from simulations and a phenomenological electrostatic theory of confined liquids.\cite{monet2021nonlocal}

Further, we explore what happens when a pure polar fluid with bulk dielectric constant $\epsilon_r=80$ is subjected to a strong electric field at the boundaries, driving the system to nonlinear response with experimentally feasible surface charge densities. Fig. \ref{fig:F_qs_var2} includes the normal dielectric permittivity, the tangential permittivity, the dipole orientation, and the dipole density as a function of distance from the left surface for different values of surface charge density. The overscreening structure and singularities in the normal dielectric constant are more or less unchanged as the surface charge increases. However, at the largest charge density, $q_s=0.25$ C/m$^2$, the normal and tangential components of the permittivity saturate to a lower value. Even though the strong electric field leads to dielectric saturation and electrostriction (an increase in the local dipolar concentration near the interface), the saturation of the orientation of the dipoles due to the strong field intensity leads to a lower effective dielectric constant for the polar fluid in the nanoslit. Interestingly, the dipolar concentration profile at high charge density forms layers with sharp cusp-like peaks near the surface.

Finally, the behavior of the polar liquid model is studied as a function of the extent of confinement, as shown in Fig. \ref{fig:F_cos_theta_L}. Here, the distance between the two surfaces, $L$, is varied between $L=0.36$ nm to $L=1.08$ nm. The layering in the orientation of dipole molecules relative to the normal axis, $\langle \cos(\theta)\rangle$, is complicated by the coherency of the layers of charge emanating from each surface. As the surface separation increases, the dipoles go from one layer, to two layers, to four layers, back to three layers, then three layers with smaller sublayers, then four layers. This means that single angstrom differences in separation can lead to constructive or destructive interference from opposing layers of dipoles, forming dipolar patterns with varying orientations, periodicity, and number of layers.

\begin{figure*}
\centering
\includegraphics[width=1 \linewidth]{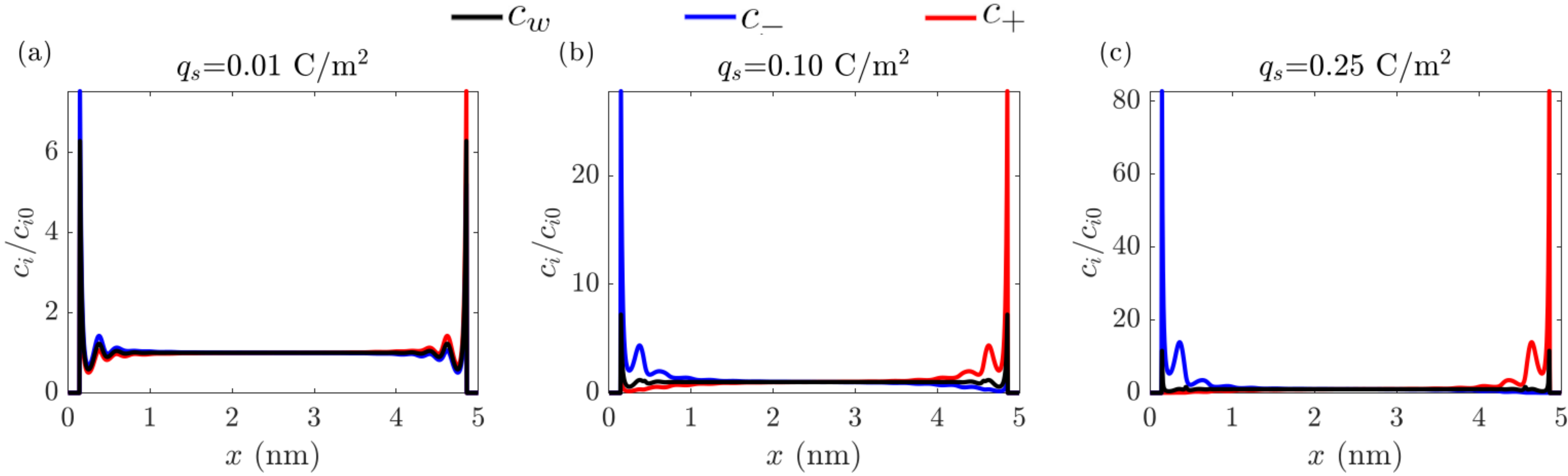}
\caption{Electrolyte screening behavior between two surfaces of opposite charge for varying surface charge density. The curves are generated by solving Eq. \ref{eq:1D_form} with $\bar{\rho}_e \neq 0$. The surface charge density is varied between $q_s=0.01$ C/m$^2$,$q_s=0.1$ C/m$^2$, and $q_s=0.25$ C/m$^2$, keeping all other parameters constant ($T=300$ K, $L=5$ nm, $c_{w0}=55$ M, $d=0.285$ nm, $c_0=0.1$ M,  and $p_0=4.86$ D). The cation (red), anion (blue), and dipolar molecule (black) profiles are plotted, normalized to their respective bulk values.  }
\label{fig:F_qs_var_conc}
\end{figure*}

\subsection{What changes in the presence of a strong electrolyte}
Commonly, dissolved ions are present in polar fluids due to dissociation of electrolytes. In this section, we examine the screening of charge at interfaces for a polar fluid with nonzero ion concentration. Due to the nonzero ion concentration, the surface charges need not be equal nor do they need to be opposite, since any net charge of the two surfaces will be screened by ionic charges in the nanoslit. For the purposes of this section, however, to stay in line with the pure polar fluid case, we maintain equal but opposite surface charge density on the bounding walls of the nanoslit.

In Fig. \ref{fig:F_conc_var}, we highlight the main electrostatic properties for a 1:1 electrolyte of varying ionic concentration, $c_0$. Examining the potential, $\phi$, in Fig. \ref{fig:F_conc_var}(a), the overscreening oscillations seem to only weakly depend on the ionic concentration. The difference in the local concentrations of anions and cations, $c_--c_+$, includes oscillatory structures for all concentrations, owing to the dielectric overscreening. Further, at the highest concentration, the ions themselves also contribute to overscreening, where the local ionic charge density oscillates between negative and positive values. While the ionic concentration has a weak influence on the tangential dielectric permittivity, it strongly influences the normal component of the effective solvent dielectric permittivity. The nonzero ionic concentrations lead to longer range oscillations and more singularities in the normal component of the dielectric permittivity of the solvent, ${\epsilon}_\perp$. For the highest concentration, the apparent normal dielectric permittivity appears to enter an exotic region $0<{\epsilon}_\perp<1$,  which is of course forbidden for the general dielectric constant ${\epsilon}_\perp^*$,\cite{dolgov1981admissible} but not for the effective dielectric constant,
${\epsilon}_\perp$. Note that ${\epsilon}_\perp$ is an effective quantity and the constraint 
of a forbidden band between 0 and 1,\cite{dolgov1981admissible}  does not apply to it.
 The general dielectric function, ${\epsilon}^*_\perp$, includes both the polarization of the dipoles \textit{and} the polarization from ions. For our geometry, it is defined by:
\begin{equation}\label{eq:eps_star}
    \frac{d}{dx}\left({\epsilon}^*_\perp\frac{d\phi}{dx}\right)=0
\end{equation}
and is forbidden from the region $0<{\epsilon}^*_\perp<1$ by the stability requirement. Here, we plot both ${\epsilon}_\perp$ (Fig. \ref{fig:F_conc_var}(c)) that satisfies Eq. \ref{eq:perp_perm}, and ${\epsilon}^*_\perp$ (Fig. \ref{fig:F_conc_var}(a)) that satisfies Eq. \ref{eq:eps_star}. Therefore, the weighted ionic charge density acts as a source that allows for the effective solvent normal dielectric tensor component ${\epsilon}_\perp$ to enter the region $0<{\epsilon}_\perp<1$ when ionic overscreening occurs at high ionic concentration. 

Next, the role of surface charge density on the accumulation of ions and dipolar molecules is investigated.  Fig. \ref{fig:F_qs_var_conc} shows the anion, cation, and dipolar molecule density rescaled to their bulk values ($c_0=0.1$ M for the ions and $c_{w0}=55$ M for the dipoles) as the surface charge density is varied from $q_s=0.01$ C/m$^2$ to $q_s=0.25$ C/m$^2$ . While the interfacial dipole concentration is increased at large electric field magnitudes, the counterion concentration increases much more rapidly with increasing surface charge density, relative to the bulk concentration. This is quite natural  as electrostriction in dense polar liquids in the electric field of the electrical double layer is a much weaker effect than the compression of the double layer with increased voltage drop across it. Therefore, large surface potentials preferentially accumulate counterions instead of the dipolar molecules. But when the voltage drop is large, overcrowding of counterions can occur, where layers rich in counterions of the same sign form near each of the two surfaces, pushing out the dipolar molecules further from the surface. Clearly, this conclusion depends on the size of ions and of dipoles. In the case studied here, they are of the same size. But had this been different, for example, for a situation wheree the dipoles are much smaller than the ions, then the dipoles will be drawn into the double layer to screen the repulsions between the counterions.

\subsection{Hydration forces}
The layering of charged dipolar molecules confined between two interfaces causes an oscillatory hydration interaction. The hydration interaction is critical in colloidal stability, including describing forces experienced by charged biological proteins, lipid bilayers, or DNA at the nanometer scale. Here, we show that the dipolar shell theory can capture the oscillatory hydration forces commonly observed in the measurements of the forces between smooth surfaces separated by liquid films. 

To calculate the disjoining pressure, we can use the following definition for the electrostatic free energy:
\begin{equation} 
    \mathcal{F}^\mathrm{el}[ \phi] =  \int d\mathbf{r} \Big\{\frac{\epsilon_0}{2}(\nabla\phi)^2\Big\}.
\end{equation}
and the forms of $\mathcal{F}^\mathrm{id}$ and $\mathcal{F}^\mathrm{ex}$ in Eqs. \ref{eq:eq_Fid} and \ref{eq:eq_Fex} to compute the overall free energy as a function of the separation distance between two surfaces. In the calculation, we assume free exchange with a bulk reservoir at fixed concentration, the grand canonical ensemble. The grand potential can be written as:
\begin{equation}
    \Omega=\mathcal{F}-\sum_i\int d\mathbf{r} \Big\{\mu_{ib} c_i\Big\}
\end{equation}
where $\mu_{ib}=\delta \mathcal{F}/\delta c_i\rvert_b$. The disjoining pressure can be calculated using the relation:
\begin{equation}
    P=-\frac{d \left(\Omega/A\right)}{d L}
\end{equation}
at constant temperature and reference chemical potential, where $A$ is the area of the surfaces.\cite{andelman2006introduction, ravikovitch2006density, ravikovitch2006density2, henderson2005statistical} Here, we numerically compute the integrals that define $\Omega/A$ at various values of $L$, and numerically take the derivative to arrive at the pressure. Pressures are reported relative to the bulk reference value as $L\rightarrow \infty$, $P_\infty$. 

\begin{figure}
\centering
\includegraphics[width=1 \linewidth]{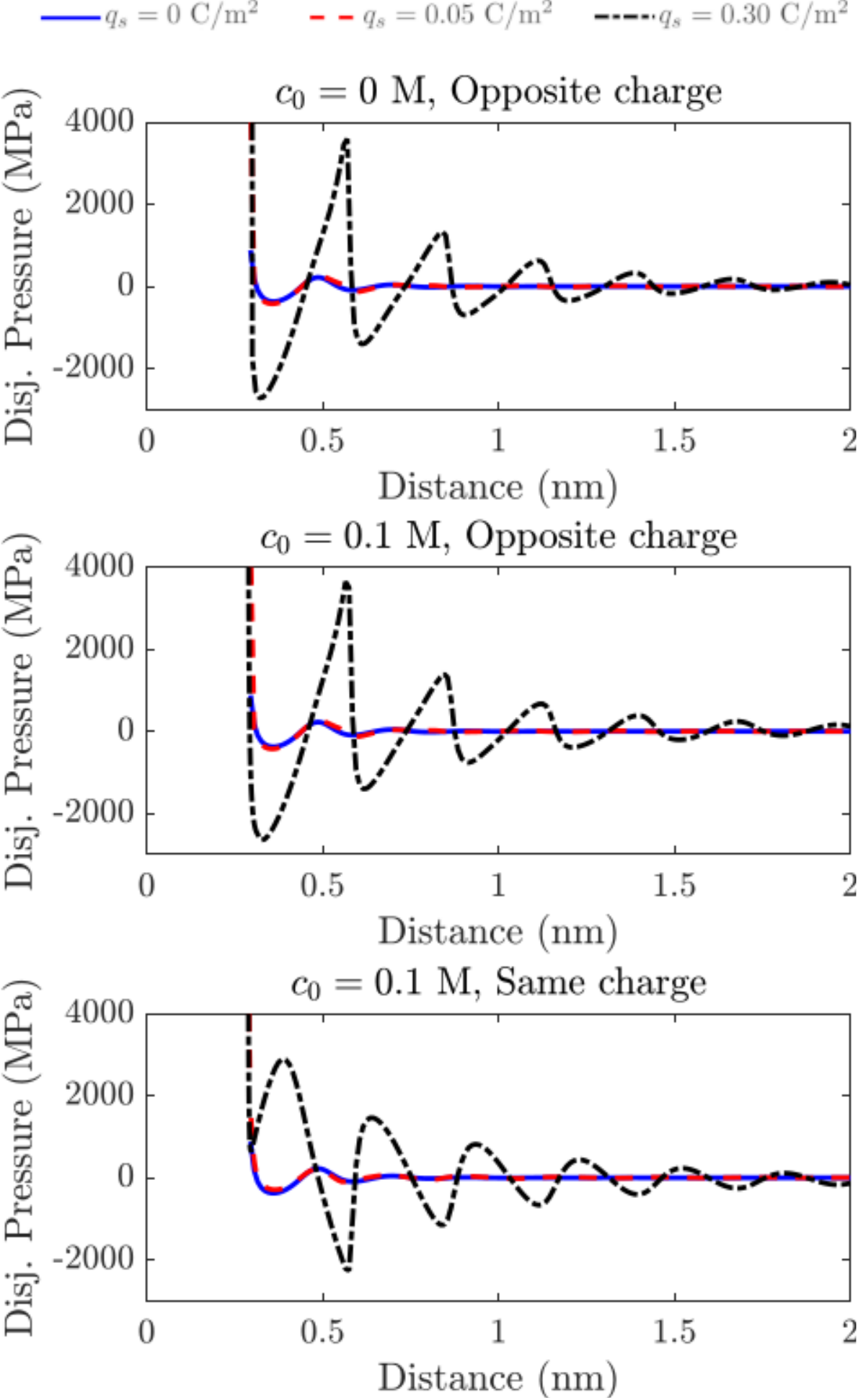}
\caption{The disjoining pressure between two surfaces of varying separation distance plotted for different ionic concentrations and polarity of surface charges. Individual curves correspond to the indicated surface charge density, held constant for all separation distances. The ionic concentration, $c_0$, and the polarity of the surfaces are listed in the figure titles. The three lines in each plot correspond to values of the surface charge density from $q_s=0$, 0.05, and 0.30 C/m$^2$.   }
\label{fig:F_surf_force_var}
\end{figure}

In Fig. \ref{fig:F_surf_force_var}, the disjoining pressure is plotted for (a) a pure polar fluid between two surfaces of opposite charge  (b) a 0.1 M 1:1 electrolyte between two surfaces of opposite charge, and (c) a 0.1 M 1:1 electrolyte between two surfaces of the same charge. First we will discuss case (a) of the pure polar fluid. At zero surface charge density, the interactions are dominated by the packing effects captured in $\bar{\mu}^\mathrm{ex}_i$. At larger charge density, such as $q_s=0.30$ C/m$^2$, the electrostatic contribution to the disjoining pressure dominates the interaction. While the initial few layers of the profile are jagged, the pressure profile gives way to regularly-shaped decaying oscillations at larger separation distances. For case  (b), adding in an electrolyte at low concentration ($c_0=0.1$ M) relative to the dipole concentration ($c_{w0}$=55 M) does not significantly change the observed patterns in the short range hydration interaction at low or high surface charge density. However, if as shown in case (c), the charge on the two surfaces is of the same sign, then the oscillation pattern is shifted and the sharpness in the patterns is flipped. Even so, the general pattern for interactions between surfaces of the same charge and of opposite charge are relatively similar in their overall envelope and long-range decay. Such an angled pressure profile dependent on the surface charge polarity, while not immediately discernible in SFA experiments in the literature, could be detected with a carefully designed experiment, if it is in fact present. Furthermore, measurements with soft surfaces or surfaces that are rough might blur these predicted features

\subsection{Model applied to double layer capacitance}
In electrochemistry, one of the important measureable interfacial quantities is the double layer capacitance. In traditional theoretical approaches, the capacitance is composed of a constant Stern capacitance, $C_s$ and a Gouy-Chapman diffuse layer capacitance $C_D$ in series. The Stern capacitance is assumed to arise due to the layer of water hydrating the interface with depressed dielectric constant and fixed thickness. The diffuse layer capacitance accounts for the screening of the surface charge by the ionic charge distribution in the solution near the interface. The total differential capacitance of an electrode, $C_T$, is defined as:
\begin{align}
    C_T=\mid \frac{d q_s}{d \phi_s} \mid,
\end{align}
where $\phi_s$ is the surface potential.  The total differential capacitance is therefore related to the Stern and Debye capacitance:
\begin{align}
    C_T=\left(C_D^{-1}+C_s^{-1}\right)^{-1}.
\end{align}
At small potential drops across the double layer for dilute solutions, the diffuse layer capacitance is approximatley equal to the Debye capacitance:
\begin{align}
    C_D=\frac{\epsilon_r\epsilon_0}{\lambda_D}
\end{align}
where $\lambda_D$ is the Debye length,
\begin{align}
    \lambda_D=\sqrt{\frac{\epsilon_r \epsilon_0 k_B T}{2e^2 c_0}}.
\end{align}
In the dipolar shell theory, the equally-sized hard sphere assumption means that there is no layer of water near the surface. In the model, the water layering and ionic screening occur in a diffuse manner from the interface. Despite the overlap, the layering of water still leads to an \textit{effective} Stern capacitance.\cite{kornyshev1982nonlocal} 

\begin{figure}
\centering
\includegraphics[width=1 \linewidth]{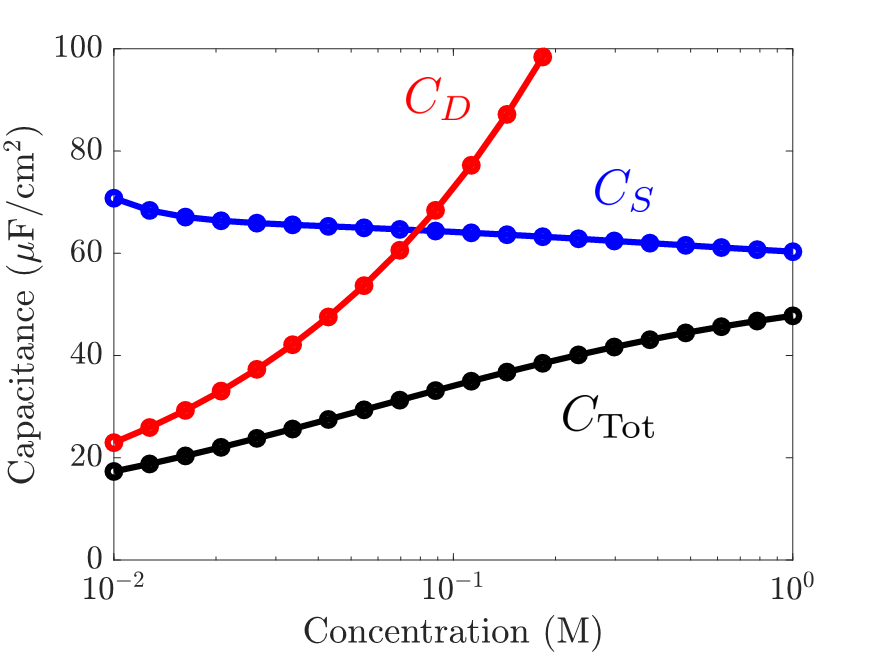}
\caption{Effective capacitance at zero charge for an electrolyte as a function of ionic concentration. All other parameters are kept constant  ($T=300$ K, $L=5$ nm, $c_{w0}=55$ M, $d=0.285$ nm,  and $p_0=4.86$ D). The total capacitance is calculated numerically from an isolated (non-overlapping) double layer. The Debye capacitance is calculated as $C_D=\epsilon_r\epsilon_0/\lambda_D$, and the effective Stern capacitance is calculated assuming a series capacitance model to match the total capacitance from the dipolar shell theory. }
\label{fig:F_capac}
\end{figure}

In Fig. \ref{fig:F_capac}, we calculate the variation in the total double layer capacitance, effective Stern capacitance, and calculated Debye capacitance  as the ionic concentration changes, all calculated near the point of zero charge for non-overlapping double layers. While the theory does not contain a specific layer of water at an interface like the traditional Stern layer concept, it returns a nearly-constant effective Stern capacitance around $60$ $\mu$F/cm$^2$. The general predictions of semi-phenomenological nonlocal electrostatic theory in ref. \cite{kornyshev1982nonlocal} is fully supported by this ``molecular" level model. The details of the capacitance could be affected strongly by the size asymmetry of real polar liquids and ions. In other words, small water molecules would access the surface more easily than larger ions in the solution, in order to reduce the electrostatic repulsion between the counterions. Furthermore, the induced polarization of the solvent and ions can strongly affect the capacitance.\cite{hatlo2012electric}

\subsection{Linearized form of equations}
The system of equations outlined above are, generally, nonlinear integro-differential equations. While they can be solved in a straightforward manner numerically, they do not admit simple analytical solutions. Here, we show how the system of equations can be reduced to linear differential forms, where the oscillatory decay can be described analytically. While not valid for the first few layers of oscillations in charge and mass, the linearized forms of the theory are decent approximations for the long-range behavior of the polar fluid.

The linearized weighted Langevin-Poisson equation for the dipolar shell theory (combining linearized forms of Eqs. \ref{eq:eqP_bar_def}, \ref{eq:eqIonicCharge}, and \ref{eq:eq_A}) is:
\begin{align}
    \left[1+(\epsilon_r-1)\hat{w}_s^2\right]\nabla^2\phi=\epsilon_r \kappa_D^2 \hat{w}_s^2\phi
\end{align}
where $\kappa_D$ is the inverse Debye length and $\phi$ is written in its local form. For small perturbations, we can assume that the convolution with $w_s$ acts as a differential operator, $\hat{w}_s\approx1+\ell_s^2\nabla^2$, where $\ell_s=d/\sqrt{24}$.\cite{desouza2020interfacial} The $\hat{}$ symbol correpsonds to the differential form of the weighting function. In $1D$, if we assume $\phi=A\exp(\kappa x)$, we get the following characteristic equation for the decaying modes:
\begin{align}
    \kappa^2+(1+\kappa^2\ell_s^2)^2\left[(\epsilon_r-1)\kappa^2-\epsilon_r\kappa_D^2\right]=0.
\end{align}
If there is no electrolyte present, $\kappa_D=0$, then the solution for $\kappa$ has a real and imaginary part as:
\begin{equation}
    \begin{split}
        \mathrm{Re}(\kappa)&=0,\quad \pm \frac{1}{\ell_s}\sqrt{-\frac{1}{2}+\frac{1}{2}\sqrt{\frac{\epsilon_r}{\epsilon_r-1}}} \\
        \mathrm{Im}(\kappa)&=0,\quad \pm \frac{1}{\ell_s}\sqrt{\frac{1}{2}+\frac{1}{2}\sqrt{\frac{\epsilon_r}{\epsilon_r-1}}}.
    \end{split}
\end{equation}
In the limit of large $(\epsilon_r-1)$, we get:
\begin{equation}
    \begin{split}
        \mathrm{Re}(\kappa)&=0,\quad \pm \frac{1}{2\ell_s\sqrt{\epsilon_r-1}} \\
        \mathrm{Im}(\kappa)&=0,\quad \pm \frac{1}{\ell_s}.
    \end{split}
\end{equation}

This means that in the absence of ions, the effective hydration length governing decay of the oscillations scales as:
\begin{equation}\label{eq:eqlam_s}
    \lambda_s=d\sqrt{(\epsilon_r-1)/6},
\end{equation}
and an oscillation wavelength of about one molecular diameter. For water, the effective hydration length turns out to be $\approx 1$ nm at room temperature. The hydration length describes the decay of the alternating layers of bound charge emanating from a surface, and the magnitude of this decaying mode is determined by the magnitude of polarization at the surface.

Now, if we include salt, the expressions for the decaying modes become more complicated. In the limit of small but non-zero ion concentrations, $\kappa_D\rightarrow 0$ gives a longest decaying mode of of $\kappa=\kappa_D$, where the effective decay length is the Debye length, $\lambda_D$. In this limit, the hydration length can be thought to be independent of and additive to the long range Debye screening, as is commonly assumed in experimental measurements of surface forces.\cite{kornyshev1983non} 

In the limit of large ionic concentrations and large $(\epsilon_r-1)$, another simplified formula can be attained for the slowest decaying mode:
\begin{align}
    \kappa\approx \frac{1}{2 \sqrt{\epsilon_r} \kappa_D \ell_s^2}\pm\frac{i}{\ell_s}.
\end{align}
This formula is valid when $\sqrt{\epsilon_r}\kappa_D \ell_s\gg 1$, meaning that the molecule size is much larger than the Debye length in vacuum. The effective screening length becomes independent of the relative permittivity,
\begin{equation}
    \lambda_s\approx \frac{2\ell_s^2\sqrt{\epsilon_r}}{\lambda_D}=\frac{d^2\sqrt{\epsilon_r}}{12\lambda_D}
\end{equation}
since the dependence of $\epsilon_r$ cancels out, and the oscillations are on the order of one molecular diameter. 

An additional source of oscillations is from the density variations owing to packing of molecules at a flat interface. Taking each set of species, we can assume small perturbations:
\begin{equation}
    \begin{split}
        &c_w\approx c_{w0}\left(1-\beta\bar{\mu}^\mathrm{ex}\right)\\
        &c_i\approx c_{i_0}\left(1-z_ie\beta \bar{\phi}-\beta\bar{\mu}^\mathrm{ex}\right).
    \end{split}
\end{equation}
 If we sum over all species assuming the same size of each molecule and thus identical excess chemical potential, we get an expression for the local filling fraction:
\begin{equation}
    \eta-\eta_0=-\eta_0\beta\bar{\mu}^\mathrm{ex}=-\frac{2\eta_0(4-\eta_0)}{(\eta_0-1)^4}\hat{w}_v^2\left(\eta-\eta_0\right)
\end{equation}
where we have linearized the excess chemical potential. We can again assume small perturbations and treat $w_v$ as an operator, $\hat{w}_v\approx 1+\ell_v^2\nabla^2$, where $\ell_v=d/\sqrt{40}$. The decay of the density for a symmetric fluid is independent of the decay of the electrostatic potential, and the characteristic equation, assuming $\eta-\eta_0=A\exp(-\kappa_m x)$, is:
\begin{align}
    1+\frac{2\eta_0(4-\eta_0)}{(\eta_0-1)^4}(1+\ell_v^2\kappa_m^2)^2=0.
\end{align}
Here, $\kappa_m$ has real and imaginary parts:
\begin{equation}
    \begin{split}
        \mathrm{Re}(\kappa_m)&=\frac{1}{\ell_v}\sqrt{-\frac{1}{2}+\frac{1}{2}\sqrt{1+\frac{(1-\eta_0)^4}{2\eta_0 (4-\eta_0)}}}\\
        \mathrm{Im}(\kappa_m)&=\frac{1}{\ell_v}\sqrt{\frac{1}{2}+\frac{1}{2}\sqrt{1+\frac{(1-\eta_0)^4}{2\eta_0 (4-\eta_0)}}}.
    \end{split}
\end{equation}
As $\eta_0\rightarrow 0$, the oscillations decay rapidly over a small length scale $\lambda_m=\ell_v(8\eta_0)^{1/4}$. For dense solutions, as $\eta_0\rightarrow 1$, the real and imaginary parts of the solution go to:
\begin{equation}
    \begin{split}
        \mathrm{Re}(\kappa_m)&=\frac{(1-\eta_0)^2}{\sqrt{8}\ell_v\sqrt{\eta_0 (4-\eta_0)}}\\
        \mathrm{Im}(\kappa_m)&=\frac{1}{\ell_v}.
    \end{split}
\end{equation}
In other words, the decay of mass oscillations with wavelength of the molecular diameter goes as:
\begin{align}\label{eq:eqlam_m}
    \lambda_m\approx \sqrt{\frac{1}{5}}\frac{d\sqrt{\eta_0 (4-\eta_0)}}{(1-\eta_0)^2}.
\end{align}
For water at room temperature, the decay length for mass oscillations is $\lambda_m\approx 0.4$ nm. Therefore, the mass density oscillations decay more rapidly (over a shorter length scale) than the oscillations in the potential. While the typical filling fraction for pure liquids is around $\eta_0\approx 0.4$, an increase in the filling fraction corresponds to longer range oscillations in the liquid number density. The competition between mass and hydration length depends on the filling fraction of the fluid and the relative dielectric constant of the fluid, as well as the magnitude of the surface charge density or surface potential. Even so, the oscillation wavelength for electrostatics and density variations remains comparable to the molecular or ionic diameter for each decay mode.

Thus, for highly charged surfaces, we expect to see oscillation patterns with the period of oscillations of the liquid molecule size and decay envelope of the order of 1 nm determined by the hydration length in Eq. \ref{eq:eqlam_s}, whereas for low charged or uncharged surfaces, the envelope will be much shorter based on Eq. \ref{eq:eqlam_m}. In experiments, people saw a  variance of decay length of the force between neutral surfaces,\cite{leikin1993hydration}  but this was explained by lateral inhomogeneity of the generally electroneutral charge distribution along the surface.\cite{leikin1990theory} The oscillations themselves, get smeared  by the smearing of the surfaces. Our model suggests that the interplay of charge ordering and packing effects will influence the observed decaying modes, as was demonstrated clearly by studying the hydration forces in Fig. \ref{fig:F_surf_force_var} with varying surface charge magnitudes.

\section{Conclusions}
The dipolar shell theory describes layering in charge and mass for an interfacial polar fluid. In this work, we have demonstrated that the effective delocalized bound charge on the dipolar molecules underlies the overscreening phenomenon, with alternating layers of bound charge density on the dipoles.

The overscreening effect leads to significant anisotropy in the normal and tangential components of the permittivity. The normal component has singularities owing to the overscreening effect, while the tangential component scales more closely with the dipole concentration. The length scale governing the decay of oscillations from the interface is the \textit{hydration length}, $\lambda_s=d\sqrt{(\epsilon_r-1)/6}$.

When ions are present, the ionic layering is influenced by the structuring of the polar fluid, and the ions also begin to contribute to the overscreening effect when they reach a sufficiently high concentration. 

The theory could be extended further and applied to various other applications not mentioned in this work. Straightforward extensions of the theory could describe: (i) Varying electrolyte composition with multivalent ions and mixtures of polar fluids, (ii) varying the geometry of the pore domain to cylindrical or spherical pores or using the theory to describe the double layer structure around cylindrical or spherical charged colloids, (iii) extending the analysis to non-uniform ion and water sizes,  (iv) demonstrating the charging dynamics of the dipolar fluid orientation and layering, (v) showing the role of double layer and hydration oscillation overlap on the system capacitance, (vi) demonstrating the role of the dipolar fluid organization on the effective $\zeta$-potential for electrokinetic measurements, and (vii) including the interfacial polar liquid structure in a formulation of interfacial electrochemical reactions.    

Although the theory does capture the charge structuring at the interface, it still falls short of perfectly describing real polar liquids. For example, the theory does not reproduce the single-ion-level hydration. A more sophisticated approach may be necessary to keep track of the bound and free states of water that constitute the coordinated hydration shell of individual ions. Furthermore, the theory only indirectly captures the correlations between neighboring dipolar molecules, which requires an effective dipole moment that is larger than the true value for highly polar fluids.  

The dipolar shell structure assumed in the theory is significantly simpler than typical charge distributions within polar molecules. For example, the higher-order multipole moments of water can strongly influence the interfacial polarization. \cite{bonthuis2012profile, gongadze2013quadrupole, slavchov2014quadrupole} The model assumes a hard sphere repulsion, but real polar fluids will have softer repulsive interactions, as well as attractive dispersion interactions. Finally, we have neglected electronic degrees of freedom, always present in polar fluids, which contribute to the dielectric constant of the liquid independent of the fixed dipole orientations.

Despite the simplifications, the dipolar shell theory presents a powerful theoretical framework to investigate the interfacial properties of polar liquids. The system of integro-differential equations is readily soluble, especially in 1D geometries. The approximate, differential-form of the theory gives analytical formulas for quick computations and experimental comparisons.

\section{Acknowledgements}
 All authors acknowledge the support from the MIT-Imperial College Seed Fund. JPD and MZB acknowledge support from the Center for Enhanced Nanofluidic Transport, an Energy Frontier Research Center funded by the U.S. Department of Energy, Office of Science, Basic Energy Sciences under Award \# DE-SC0019112. JPD also acknowledges support from the National Science Foundation Graduate Research Fellowship under award number \#1122374. A.A.K would like to acknowledge the research grant by the Leverhulme Trust (RPG-2016- 223). We thank Karina Pivnic, Aditya Limaye, and Michael Urbakh for useful discussions.

\bibliography{REF.bib}
\appendix
\section{Derivation of electrostatic free energy}
Here we reduce the electrostatic free energy to the form presented in the main text. The free energy can be defined as:

\begin{equation}\label{eq:Fel_eq}
    \mathcal{F}^\mathrm{el}[ \phi] =  \int d\mathbf{r} \Big\{\frac{\epsilon_0}{2}(\nabla\phi)^2\Big\}.
\end{equation}

The modified form of the Poisson's equation can be written as:
\begin{equation}
    0=\epsilon_0 \nabla^2\phi+\bar{\rho}_b+\bar{\rho}_e.
\end{equation}
in terms of the weighted bound charge density or as:
\begin{equation}
    0=\epsilon_0 \nabla^2\phi-\nabla\cdot\mathbf{\bar{P}}+\bar{\rho}_e.
\end{equation}
in terms of the weighted polarization vector.

Here, we employ a Lagrange multiplier, $\lambda$, to enforce the modified Poisson equation in our minimization of the electrostatic free energy:
\begin{equation}
    \mathcal{F}^\mathrm{el}[ \phi, \bar{\rho}_e, \mathbf{\bar{P}}, \lambda] =  \int d\mathbf{r} \Big\{\frac{\epsilon_0}{2}(\nabla\phi)^2+\lambda\left(\epsilon_0 \nabla^2\phi-\nabla\cdot\mathbf{\bar{P}}+\bar{\rho}_e\right)\Big\}.
\end{equation}
Taking a variation with respect to $\phi$, we find that $\lambda=\phi$. Plugging in this dependence gives:
\begin{equation}
    \mathcal{F}^\mathrm{el}[\phi, \bar{\rho}_e,\mathbf{\bar{P}}] =  \int d\mathbf{r} \Big\{\frac{\epsilon_0}{2}(\nabla\phi)^2+\phi\left(\epsilon_0 \nabla^2\phi-\nabla\cdot\mathbf{\bar{P}}+\bar{\rho}_e\right)\Big\}.
\end{equation}
Using the divergence theorem, we can find the following identity:
\begin{equation}
    \begin{split}
        \phi\left(\epsilon_0\nabla^2\phi-\nabla\cdot\mathbf{\bar{P}}+\bar{\rho}_e\right)&=\\
        &\nabla\cdot(\epsilon_0 \phi \nabla\phi)-\epsilon_0 (\nabla\phi)^2\\
        &-\nabla\cdot\left(\phi\mathbf{\bar{P}}\right)+\mathbf{\bar{P}}\cdot\nabla \phi\\
        &+\bar{\rho}_e \phi.
    \end{split}
\end{equation}
Here, we can take the divergence terms to a surface far away, where the potential and related fields are zero, leaving the following expression for the electrostatic energy:
\begin{equation}
    \mathcal{F}^\mathrm{el}[{ \phi}, \bar{\rho}_e, \mathbf{\bar{P}}] =  \int d\mathbf{r} \Big\{-\frac{\epsilon_0}{2}(\nabla\phi)^2+ \bar{\rho}_e \phi+\mathbf{\bar{P}}\cdot \nabla\phi\Big\}.
\end{equation}

We can confirm that $\delta \mathcal{F}^\mathrm{el}/\delta \phi$ returns the modified Poisson equation. Note that while the functional appears non-convex, the unaltered electrostatic free energy in Eq. \ref{eq:Fel_eq} ensures convexity of the electrostatic free energy. Therefore, any function for the potential that satisfies the modified Poisson equation is also guaranteed to minimize the electrostatic free energy within the constraints.

\end{document}